\pdfoutput=1
\RequirePackage{fix-cm}
\documentclass[smallextended]{svjour3}       
\smartqed  
\usepackage{graphicx}
\usepackage[cmex10]{amsmath}
\usepackage{mathabx}
\usepackage{algorithmic}
\usepackage{array}
\usepackage{mdwmath}
\usepackage{mdwtab}
\usepackage{eqparbox}
\usepackage{xcolor}
\usepackage{url}
\usepackage{dcolumn}
\usepackage{hyperref}
\usepackage{subcaption}
\hypersetup{
    bookmarks=true,         
    unicode=false,          
    pdftoolbar=true,        
    pdfmenubar=true,        
    pdffitwindow=false,     
    pdftitle={},    
    pdfauthor={Zhengxuan Wu},     
    pdfsubject={},   
    pdfcreator={Zhengxuan Wu},   
    pdfproducer={},  
    pdfkeywords={}, 
    pdfnewwindow=true,      
    colorlinks=false,       
    linkcolor=red,          
    citecolor=green,        
    filecolor=magenta,      
    urlcolor=cyan           
}
	
\begin{document}

\title{Uncovering Political Promotion in China: A Network Analysis of Patronage Relationship in Autocracy 
}


\author{Zhengxuan Wu \and Jason Luo \and Xiyu Zhang 
}


\institute{Z. Wu \at
              Stanford University, Stanford, CA \\
              Tel.: +216-551-7046\\
              \email{wuzhengx@stanford.edu}           
}

\date{Received: date / Accepted: date}

\maketitle

\begin{abstract}
Understanding patronage networks in Chinese Bureaucracy helps us quantify promotion mechanism underlying autocratic political systems. Although there are qualitative studies analyzing political promotions, few use quantitative methods to model promotions and make inferences on the fitted mathematical model. Using publicly available datasets, we implement network analysis techniques to advance scholarly understanding of patronage networks in autocratic regimes, using the Chinese bureaucracy as an example. Using graph-based and non-graph-based features, we design three studies to examine drivers of political promotions. We find that careers of politicians are closely associated with their genders, home origins, and positions in the patronage networks. 
\end{abstract}

\section{Introduction}
\label{intro}
Interacting with others and forming connections are important skills among top executives in large companies and government institutions. Previous literature qualitatively demonstrated that informal connections help employees go around with formal constraints in large institutions \cite{xin1996guanxi}. How the patronage network will affect promotions of politicians? What features of the network play important roles in promotions? These are the questions on which we can make several different arguments. 

Analysis around political regime suffer from lacking of data. Unlike social data from social platforms, political database is quite limited. Applying data-mining methods on the publicly available data will enable researchers to conduct quantitative studies in political science.  

Previous literature mainly focused on the qualitative measurements of effects of patronage among China's political elites. They often came from limited insider sources \cite{tsou19761}. As a result, these studies often ended up with theoretical speculations and thus lack statistical inferences. Although researchers used the term \textit{network} in political research \cite{tsou19761}, only a few scholars have applied actual social network analysis (SNA) \cite{keller2016}. Studies pointed out that the term \textit{network} was used a a metaphor representing a group of people with special ties \cite{ward2011network}. The limited studies in patronage networks mainly focused on local network features and tried to use network analysis to study current state of Politburo behaviors. They did not try to use statistical tools to infer future political characteristics based on those features \cite{https://calhoun.nps.edu/handle/10945/38937}.

We aim to combine the SNA with statistical techniques, specifically with the goal of studying promotions in government institutions. Our study draw lines across network analysis, statistical analysis and political promotions in autocratic regimes. The insights will deepen our understanding of the structures of autocratic institutions, the patronage mechanism and the promotion process.

In this paper, we applied social network analysis to the Chinese Political Elite Database (CPED) dataset \cite{jiang2017making}. Our research began with the constructions of three categories of networks - home origin, overlap-based patronage and promotion-based patronage.  

Then, we used regression models to find correlations between promotions and both external and network features among Chinese politicians in our dataset. In the study 1, we applied linear regression to find correlations between the promotion history of politicians and their direct superior. In other studies, we analyzed the correlations between one's political career and different features from the network that were constructed. Our exploratory variables included basic features from biographies and advanced features from networks, such as node level features and structural level features. Finally, based on the fitted results, we interpreted the fitted parameters from those models and try to discover network features that were correlated with an politician's promotion in career.

\section{Related Works}
\label{sec:1}
\subsection{Patronage Effect}
Patronage network plays an major role in regimes around the world, especially in China \cite{evans2012embedded,macaulay2018non,jiang2017making,xin1996guanxi,keller2016,https://calhoun.nps.edu/handle/10945/38937}. Previous studies demonstrated \textit{Guanxi}, which means \textit{patronage network} in Chinese, plays a central role in promoting employees in large institutions, such as private or public-owned companies and government \cite{jiang2017making}. Likewise, researchers state that the patronage network of the formal Prime Minister of China, Jiabao Wen was worth billions \cite{wen2012}.


Similarly, studies showed that the career promotions of politicians were closely related to their relations with colleagues in the patronage network \cite{jiang2017making}. For instance, researchers in India pointed out that having close relationships with one's superiors would increase one's chance of promotions \cite{miller1992powers}. Other studies found that forming personal connections within institutions would help employees get promotions and work around constraints \cite{jiang2017making}. We believe that the effect of the patronage network on promotions could be longitudinal. In other words, the promotion trajectory of a politician could be positively associated with the promotion trajectory of his or her closely related connections in the patronage network. 

\subsection{Social Network Analysis}
Network Analysis has been used in different fields of research, including information science \cite{doi:10.1177/016555150202800601}, kinship analysis \cite{doi:10.1177/0038038588022001007}, online service recommendations \cite{Debnath:2008:FWC:1367497.1367646}, co-author citation analysis \cite{HUMMON198939}, gene-disease analysis \cite{10.1371/journal.pone.0020284} and online social network analysis \cite{Mislove:2007:MAO:1298306.1298311}. For example, researchers identified studies with similar topics by applying graphical analysis on the co-author citation network. Likewise, by applying network analysis to gene-disease network, they learned how genetic and environmental factors, such as drugs, contributed to diseases \cite{10.1371/journal.pone.0020284}. 

In social science, researcher used network analysis to study community structures and predict structural appearances by incorporating machine learning techniques \cite{doi:10.1002/asi.20591,Girvan7821}. For instance, researchers proved that certain community structures existed in the selected networks by studying the interactions between their components \cite{doi:10.1002/asi.20591}. Similarly, recent literature stated that edges between nodes in the network could be predicted by combining machine learning and network analysis in social network \cite{Girvan7821}. Current studies also showed that graphs could be auto-generated by applying generative models with networks analysis \cite{pmlr-v80-you18a}.

\subsection{Features Extracted From Data}
The promotion of a politician is determined by many external factors, such as one's work experience, genders, leadership skills, and economic growth of the city  \cite{seibert2001proactive}. For example, studies proved that there existed a gender bias in political promotions \cite{haraway2001gender}. Particularly in work-space promotions, researcher found that, although the wage differences were not huge, female candidates had significantly lower probabilities in promotions compared to the male candidates \cite{blau2007new}.

Home origin is another external feature. Interpersonal distance is the psychological term for measuring how close people are \cite{hall1990nonverbal}. Studies proved that people who came from the same country would have stronger bounds in social networks \cite{remland1995interpersonal}. In this case, we define two people to have the same home origin if they come from the same city.

\subsection{Social Network Features}
The patronage network will affect one's promotion \cite{keller2016}. Previous studies have shown that faction, for example, will affect a politician's career. There are also correlations between faction network, schooling network and home origin network \cite{keller2016}. We will extract different network features from all the networks we built, and test their correlations with politician's trajectories. It will have node level features, including degrees, n-hop features, and structural level features, including role and motif detection.

\section{Dataset}
We use the Chinese Political Elite Database (CPED), a large biographical database that contains extensive demographic and career information of over 4,000 key city, provincial and national leaders in China since late 1990s \cite{jiang2017making}.

For each leader, the database provides information about the time, place, organization, and rank of every job assignment listed in one's curriculum vitae, which is collected from government websites, yearbooks, and other trustworthy Internet sources. The author matches each city-year spell in the panel data set with a city secretary and a mayor. In cases where multiple leaders held the same post within a given spell, the person with the latest entry date is chosen.
\begin{figure}[h]
\includegraphics[width=8cm]{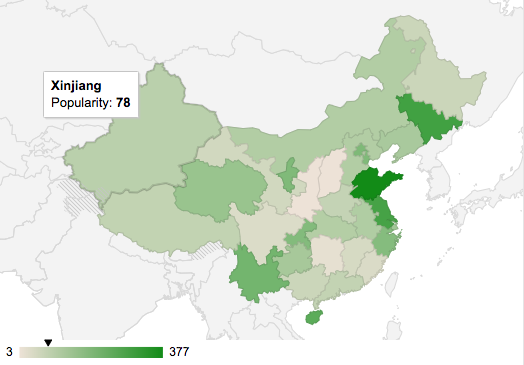}
\centering
\caption{Heatmap of hometown origin of all politicians in the dataset: Xinjiang has 78 politicians.}
\end{figure}

Figure 1 shows the distribution of politicians in the dataset according to their home origins. According to the results, we can see that the politicians are distributed across the entire country. On the other hand, most of them are from the east coast, whereas handful of politicians come from the middle of China. This maps with the population distribution in China. Table 1 shows details about the dataset, including the counts of politician, province and city. It contains the total count of data points in our work experience table. 
\begin{table}[h!]  \centering 
  \caption{Detials about the politician dataset.}
\begin{tabular}{||c c||} 
 \hline
 \textbf{Item} & \textbf{Count} \\ [0.5ex] 
 \hline\hline
 Politician & 4057 \\ 
 Province & 32 \\
 City & 389 \\
 Work Experience Data Points & 62742 \\[1ex] 
 \hline
\end{tabular}
\label{table:1}
\end{table}

\section{Network Definition}
This section describes how we use the plain text dataset of Chinese politicians to make our networks and graphs. We constructed three main networks to simulate the patronage network in real life, including hometown origin network, work experience overlap network and promotion network. The dataset included 4057 politicians in total, which were considered in all three graphs as nodes.

\subsection{Home Origin Patronage Networks}
The hometown network is highly clustered based where they come from. For each city group, all nodes are completely connected. They all have the same degrees.

\begin{figure}
    \centering
    \begin{subfigure}[b]{0.5\textwidth}
        \includegraphics[width=\textwidth]{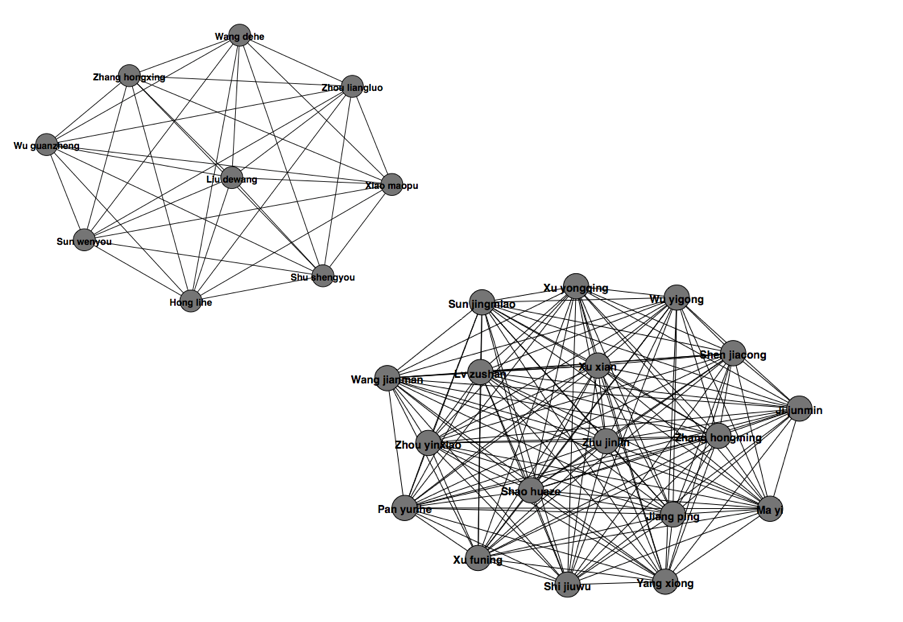}
        \label{fig:gull}
    \end{subfigure}%
    ~ 
    \begin{subfigure}[b]{0.5\textwidth}
        \includegraphics[width=\textwidth]{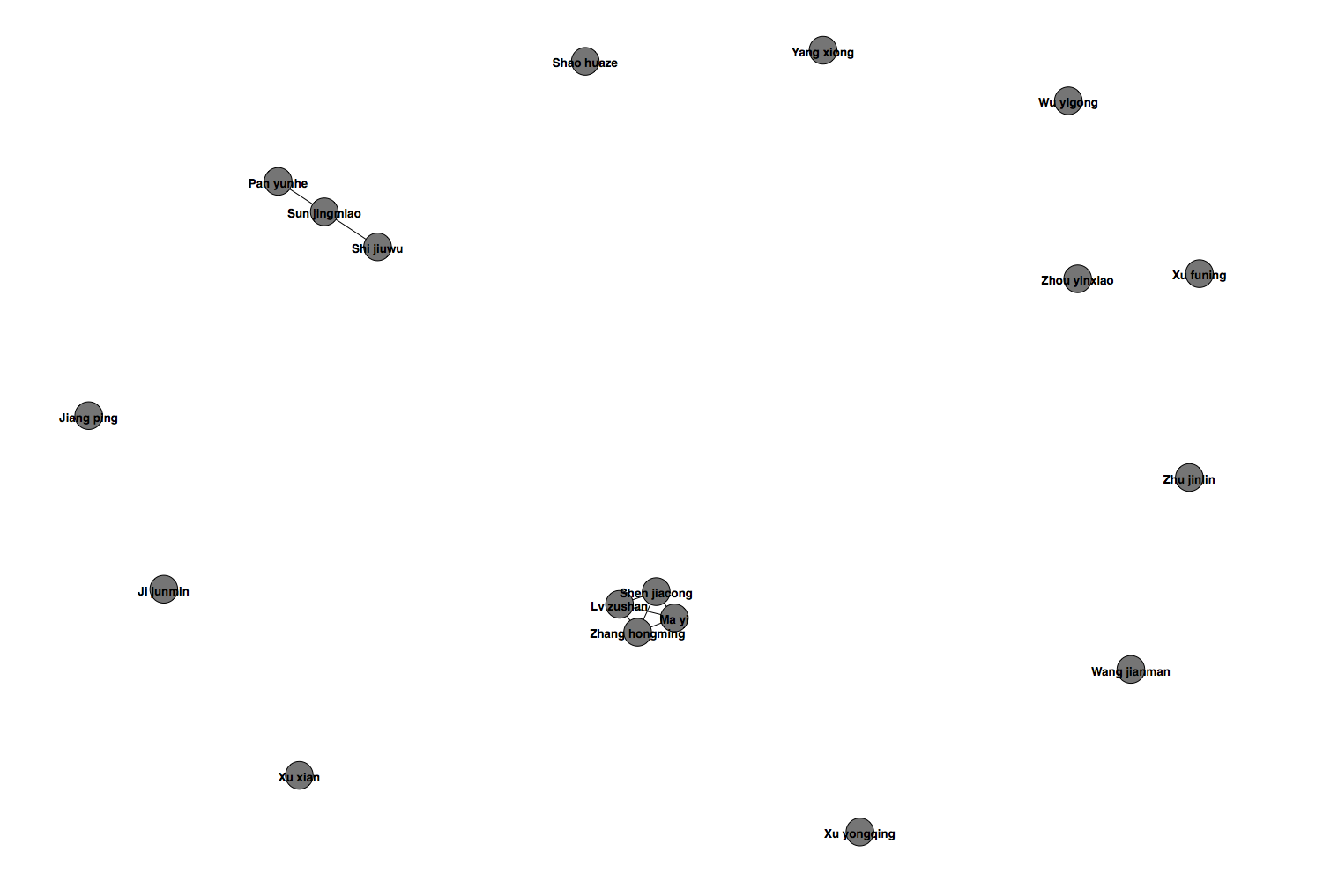}
        \label{fig:tiger}
    \end{subfigure}
    \caption{Left graph is the networks of politicians based on their home origins. This network is for all the politicians from two cities, Hangzhou(bottom right) and Shangrao(top left). Right graph is the networks of politicians based on their home origins and work experience. All these politicians are from Hangzhou and share work experience.}\label{fig:animals}
\end{figure}


We have 59306 edges within the network. Each edge represents that two politicians come from the same city. We have two different networks constructed. For the first graph, two nodes are connected if they are from the same city. For the second graph, two nodes are connected if they are from the same city and they have worked together in the same department at some point in their lives.
To avoid unreadable visualization, we plot hometown networks for two cities as an example, shown in Figure 2. All the nodes within the clusters are all inter-connected. The network on the right is less dense as we restricted the network to only have an edge between those who have worked together. As a result, many nodes are not connected. Some groups form strong connections, as those politicians are from the same city and also share working experiences.
\subsection{Overlap-based Patronage Network}
We next constructed a directed, weighted graph based on overlapping work and school experiences of the 4,057 political leaders. 
We have 655,769 directed, weighted edges. Each edge in the network indicates the existence of at least one overlapping work or school experience between the two leaders. We consider an overlap if two have to work together for at least six months in the same municipality in the same province. if multiple overlaps are founded, we encoded into to edge weights, given by the total time of two working together. The direction of an edge is determined by comparing which cadre is senior to the other in terms of their average cadre level during the time periods they work together.
The node distribution plot in Figure 3 suggests that most nodes have degrees between 100 and 1000.

\begin{figure}[h]
    \centering
    \begin{subfigure}[b]{0.5\textwidth}
        \includegraphics[width=\textwidth]{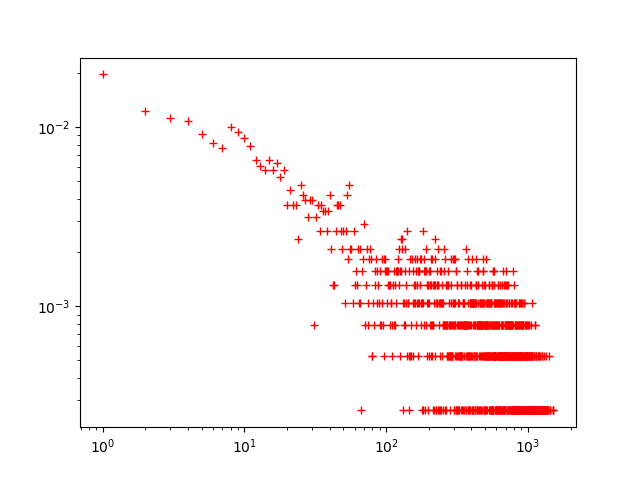}
        \label{fig:gull}
    \end{subfigure}%
    ~ 
    \begin{subfigure}[b]{0.5\textwidth}
        \includegraphics[width=\textwidth]{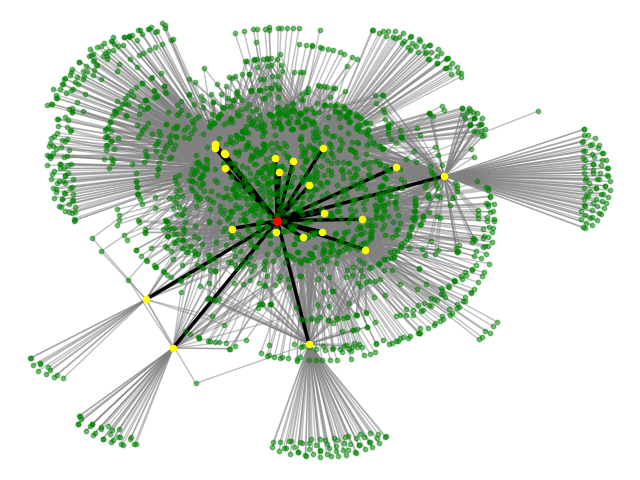}
        \label{fig:tiger}
    \end{subfigure}
    \caption{Left graph is node Degree Distribution. Right graph is two-hop Neighbors Sub-graph of A Random Node 3507.}\label{fig:animals}
\end{figure}

The right graph in the Figure 3 shows the two-hop neighbors of a random node 3057, named Zhao Zhuping, currently the head of a district in Shanghai, whose rank is equivalent to a mayor in U.S. He has 22 patronage relationships under our definition, way lower than average in our data-set. This is partly because he is relatively young and serves only a moderate position in the bureaucratic system. 

We plot node similarity graph between this node 3507 and all others to see how typical this node is and what roles other node have.
\begin{figure}[h]
\centering
  \centering
  \includegraphics[width=8cm]{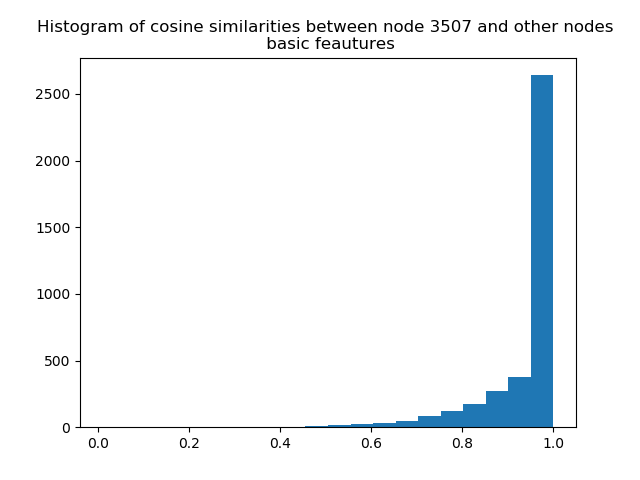}
\caption{The Graph is the node similarity distribution between Node 3507 and All Other Nodes.}
\label{fig:test}
\end{figure}
The plots in Figure 4 show that most of the nodes are almost identical to node 3507, possibly because the network is so dense that the basic feature aggregation was able to capture the entire graph. We also generated one-hop and two-hop similarity distributions. These two plots also show very clearly that the vast majority of nodes are  similar, which is suggesting the sub-graph we drew earlier for Node 3507 is highly representative across all nodes in spite of their differences in node degrees.
\subsection{Promotion-based Patronage Network}
Our third graph models patronage relationship network using political appointments among leaders. 


There are in total 3905 directed edges in the network. An edge exists between two nodes when there exists a patron-client link. A patron-client link from A to B forms when A is promoted by B. In our case, two nodes are connected if one node is promoted by another. Specifically, we look at promotion from rank level 4 to 5 based on the CPED dataset. The promotion between level 4 and 5 is considered as a milestone in one's political career. \cite{jiang2017making}. The edge goes from the node being promoted to the its promoter. This network is not weighted because we are only considering one promotion. 
\begin{figure}[h]
\includegraphics[width=4cm]{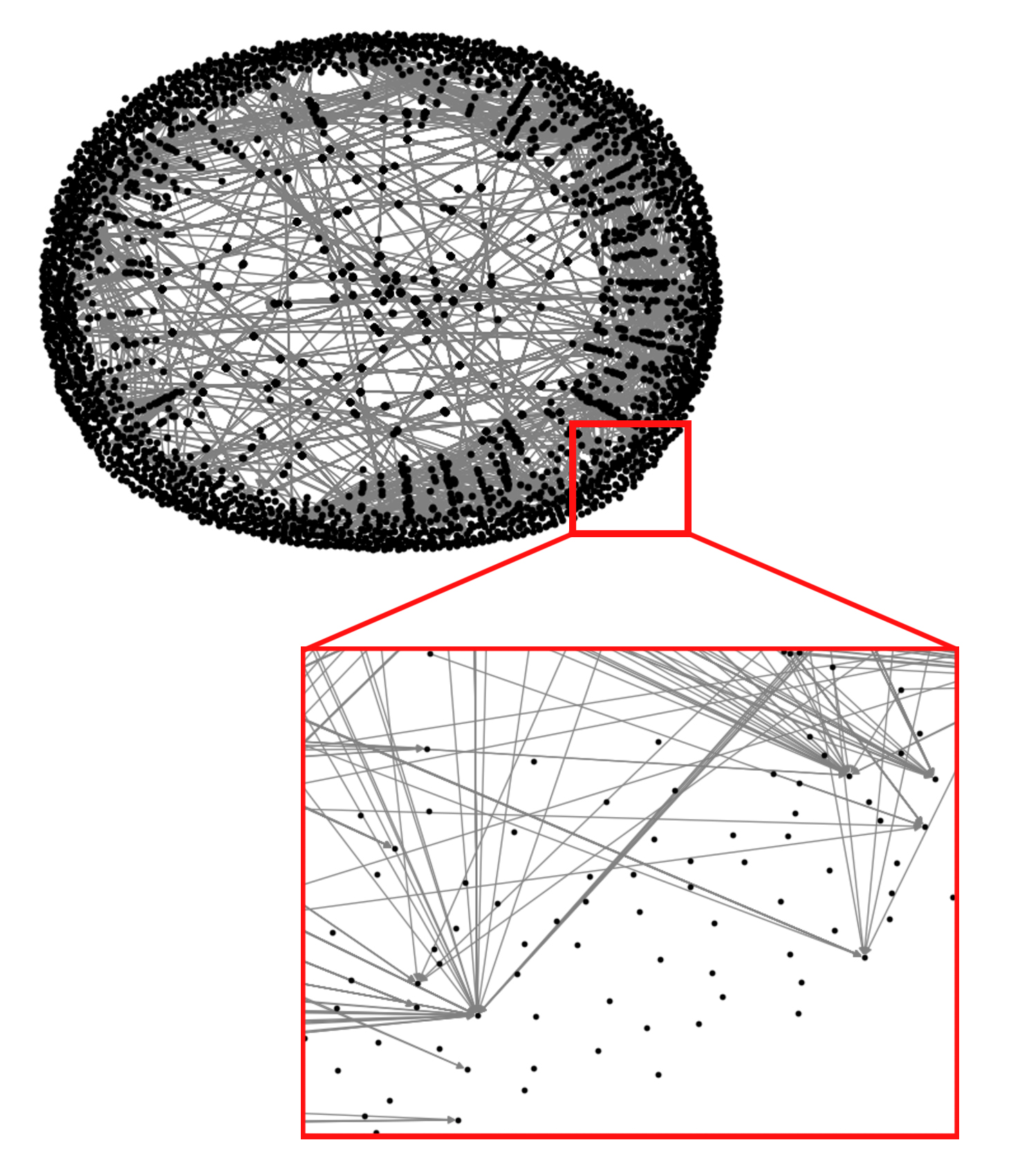}
\centering
\caption{Visualization of entire patronage network on the top, with a subset of it zoomed in to show clear edges between nodes.}
\end{figure}
\begin{figure}[h]
\includegraphics[width=8cm]{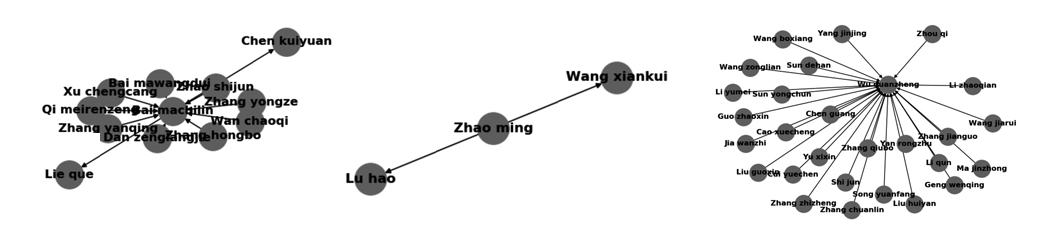}
\centering
\caption{Subgraphs with randomly chosen nodes and their edges.}
\end{figure}
To construct Figure 6, we randomly selected 30 nodes and drew out their egonets. We included all of the node's one-hop neighbors. Based on the outputs, we observe three most common structures, as shown in Figure 6.

The left subgraph shows that the node has two outgoing edges and many incoming edges. In our graph, a node can have at most two out-neighbors, because the dataset is constructed in a way that a leader can have at most two direct promoters. The node in the middle subgraph has no incoming edges, possibly because it has not reached a level that can promote others. Another common structure is shown in the right subgraph, namely, the node has only incoming edges but it has a great number of them.
\begin{figure}[h]
\centering
  \centering
  \includegraphics[width=9cm]{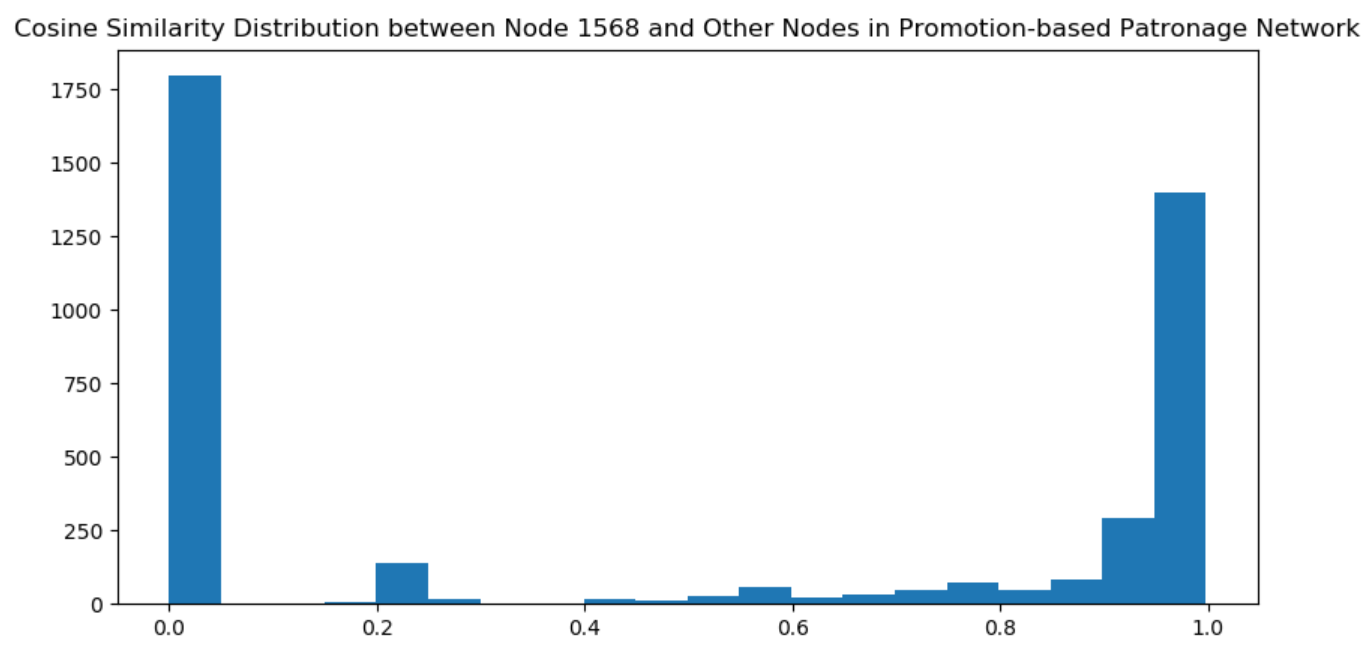}
  \label{fig:sub1}
\caption{Left graph is the node similarity distribution between Node 1568 and All Other Nodes (Two-hop features).}
\label{fig:test}
\end{figure}
As shown in Figure 7, the cosine similarity between node 1568 and other nodes reaches spikes between 0.0 and 0.5, and 0.95 and 1.0, meaning that the most of the nodes are either identical to this node or completely different from this node. There are approximately 1400 identical nodes.



\section{Method}
\subsection{Study 1: Predicting Cadre Final Rank Using Network Features}
Beyond descriptive inference, our first attempt for predictive analysis is to look at the extent to which network features predict final career result for leaders. Our baseline model is as simple as follows:
\begin{equation} \label{eq1}
\begin{split}
\text{Rank } \propto \text{ }\alpha+\beta V_{NodeFeatures}+\gamma C_{covariates}+\epsilon
\end{split}
\end{equation}
The outcome final rank is a leader's cadre level, which is one's political level in Chinese government, at 2015. For a node feature vector, we use node's feature vector up to two-hop aggregation, which is a vector of length 27. We control for one's birth year, year of joining the communist party, and year of promotion to municipality-level, which is rank 5, to adjust making comparisons across different stages of cadres' life and career.

\subsection{Study 2: Detecting Political Factions}
For this study, we analyze the effects of gender and home origin on political factions. We use all three networks in this study. For the overlap-based patronage network, we down-grade the network by removing the edge weights and directions. We assume that if two politicians have worked together, they are closely related.
\subsubsection{Gender Effect} 
For each network, we define the \textit{total degree} of each nodes as the sum of in-going edges and out-going edges. 
\begin{equation} \label{eq1}
\begin{split}
D_{i} & = \sum_{in} e(j,i) + \sum_{out} e(i,j)\\
\end{split}
\end{equation}
All the edges are undirected. For each graph, we define the proportion of nodes given a total degree as the number of nodes with a given total degree divided by the total number of nodes in the network. We look at gender differences in terms of the total degrees of nodes in the overlap-based patronage network and promotion-based network. 

We also define the rankings for all the politicians to be integers ranging from 1 to 9, with 9 being the highest level, the national leaders (Zhen guo ji). We define politicians with level above 7 as a \textit{high rank politicians}, or \textit{political elites}. For each node in our networks, we define a term \textit{average ranks of neighbors} as the average ranking of all 1-hop neighbors of the node. 
\begin{equation} \label{eq1}
\begin{split}
NbrRank_{u} & = \frac{1}{|Nbr(u)|}\sum_{v \in Nbr(u)} Rank(v) \\
\end{split}
\end{equation}
We look at the differences of the counts between male and female high rank politicians and compare the differences of average ranks of neighbors among male and female politicians.

\subsubsection{Home Origin Effect}
We use overlap-based patronage network to analyze home origin effects on political networks. After the network is down-graded, we embedded all nodes into vectors using $node2vec$ \cite{grover2016node2vec}.

$node2vec$ works by carrying out a number of random walks from each node in the graph, where the walks are parameterized by p and q. In order to validate that people with same home origin have stronger connections in the graph, we used the $BFS$ approach for $node2vec$ by setting the exploring parameters. More precisely, after having just traversed the edge from node node t to node v, the un-normalized transition probability of travelling from node v to a neighboring node x is given by:
\begin{equation}
    \alpha_{pq}(t,x)=
    \begin{cases}
      1, & \text{if}\ d_{tx}=0 \\
      1, & \text{if}\ d_{tx}=1 \\
      2, & \text{if}\ d_{tx}=2 \\
    \end{cases}
\end{equation}
We sample 10\% of nodes from the original graph before running $node2vec$. We calculate two lists of scores for each node: \textit{In Set Similarity Scores} and \textit{Out Set Similarity Scores}. The scores are calculated by taking the dot product of the embeded vectors of two nodes. Two nodes are considered as an \textit{In Set} pair if they are from the same province. For example, if two politicians are both from Shanghai, their similarity score, which is the dot product between two node vectors, would be added in to the \textit{In Set Similarity Scores} for that node. Then, we explore at the province level. For each province, we iterate through every politician from that province and concatenate their \textit{In Set Similarity Scores} and \textit{Out Set Similarity Scores}. We then define the average scores within the province as the average of each list as following:
\begin{equation} \label{eq1}
\begin{split}
score_{u} & = \frac{1}{|province_u|}\sum_{p \in province_u} score_{p} \\
\end{split}
\end{equation}
$province_u$ is the set of politicians in that province. $score$ can be either within or inter-province scores. We compare those two scores to validate the effect of home origin in political networks.
\subsubsection{Bridging Candidates Effect}
After assigning nodes to different groups, we define the groups as cliques. For each node, we define within-clique edge count as the count of edges that a node has, to the nodes that are within the same clique as itself. Similarly, we define the term between-clique edge count as the count of edges that a node has, connecting to the nodes that were in different cliques from itself.

We investigate the correlations between the within-clique edge count, between-clique edge count and rankings of politicians.

\subsection{Study 3: Hubs and Authorities}
In this study, we use the Hyperlink-Induced Topic Search (HITS) algorithm to explore the hubs and authorities among two of our patronage networks, namely, overlap-based and promotion-based networks.

HITS algorithm is an algorithm used to analyze web links. It defines two types of Web pages and calls them \textit{hubs} and \textit{authorities}. The \textit{authorities} web pages are usually prominent sources for a specific question or content. These pages are given high authority scores. On the other hand, the \textit{hubs} are those that link to authority pages and act as a guide to other authority pages, usually those with a high authority scores. These Web pages are given high hub scores \cite{Kleinberg:1999:HAC:345966.345982}.

From the dataset, we extract the rank that every political leader is at the \textit{end year} of this dataset. The \textit{end year} is defined as 2015 or the year they retire. We defined the rank as \textit{the final rank} of the politician. Every political leader has a final rank ranging from 0 to 9. We then calculate the hub and authority scores for both networks, and plot the scores against final ranks, trying to identify correlation. Scatter plot is used to show a general structure. For every rank, we take the average score of all political leaders of that rank and plot the average score on top of the scatter graph. Then we fit a linear line of the average scores against final ranks.

\section{Results}
\subsection{Study 1: Predicting Cadre Final Rank Using Network Features}
Our preliminary ordinary least squares (OLS) regression model yielded a $R^2$ of 0.625, suggesting 62.5\% of the variation in cadre final ranks are captured by our explanatory variables. Then we evaluated our baseline model's prediction power on in-sample and out-of-sample performances. For out-of-sample, we held out 10\% of the data, estimate model on the other 90\%, and test model on the 10\% for cross validation. Thus, our out-of-sample prediction was averaged over all 10 hold-out sessions. Accuracy was measured by the percentage of correct predictions of the final ranks of politicians.

\begin{table}[ht] \centering 
  \caption{In-Sample and Out-of-sample prediction performances of OLS regression model.}
\centering
\begin{tabular}{rr}
  \hline
 & Accuracy \\ 
  \hline
In Sample Predictions & 0.721 \\ 
Out of Sample Predictions & 0.718 \\
    \hline
\end{tabular}
\end{table}

The closeness of the accuracy results suggested that we are not over-fitting and we had a solid baseline results to start. We then fitted the data using more complex model, ordinal logistic regression (OLR).
{\scriptsize\begin{equation} \label{eq1}
\begin{split}
\small P(Y_i = j) &= \frac{exp(\tau_j - X\beta)}{1+exp(\tau_j - X\beta)}  - \frac{exp(\tau_{j-1} - X\beta)}{1+exp(\tau_{j-1} - X\beta)}\\
\end{split}
\end{equation}}
Assuming,
{\scriptsize\begin{equation*}
\begin{split}
Y &\sim \text{Multinomial(1, $\pi$\text)} \\
Y^* &= X\beta + \epsilon \\
Y^* &= Y(\tau_j) \\
\epsilon_j &\sim_{iid} \text{logistic} \\
\end{split}
\end{equation*}}
Here $Y_i$ is the rank outcome for each leader, and $j$ is from 0 to 9 (10 levels of cadre rank), and $X$ is feature vector up to two-hop aggregation and years of birth, joining party, and promotion, same as those in OLS. Results below suggest pooled OLS perform better than the multi-nomial logit model.

\begin{table}[ht] \centering 
  \caption{In-Sample and Out-of-sample prediction performances of pooled OLR regression model.}
\centering
\begin{tabular}{rr}
  \hline
 & Accuracy \\ 
  \hline
In Sample Predictions & 0.666 \\ 
Out of Sample Predictions & 0.649 \\
    \hline
\end{tabular}
\end{table}

Then, we used early-stage patronage network information to predict leaders' career outcome in the end. Specifically, we experimented with using network structures up until 2005-07-01 to predict cadre ranks ten years later, on 2015-07-01. 

Using similar specifications as above for OLS and OLR, except for holding off rank information post 2005, we obtained the the results in the following table.

\begin{table}[ht] \centering 
  \caption{In-Sample and Out-of-sample prediction performances on early-stage patronage network.}
\centering
\begin{tabular}{rr}
  \hline
 & Accuracy \\ 
  \hline
In Sample Predictions (OLS) & 0.690 \\ 
Out of Sample Predictions (OLS) & 0.683 \\
In Sample Predictions (Logit) & 0.725 \\ 
Out of Sample Predictions (Logit) & 0.720 \\

    \hline
\end{tabular}
\end{table}
Results above show that we have about 2.5 percentage points increase in accuracy from the OLS model, and almost no gain for the logit model. This suggests node features covering 2005 to 2015 add no gain to our prediction power. Network structures after 2005 are even noisier in predicting career outcome in 2015.

To make sure our test suffer less from model specification or baseline results, we further regressed outcome solely on node feature variables, without adding any other co-variants and outcome on nothing with the interception only. The results are showed below.
\begin{table}[ht] \centering 
  \caption{In-Sample and Out-of-sample prediction performances excluding other co-variants.}
\centering
\begin{tabular}{rr}
  \hline
 & Accuracy \\ 
  \hline
In Sample Predictions (OLS) & 0.581 \\ 
Out of Sample Predictions (OLS) & 0.575 \\

    \hline
\end{tabular}
\end{table}
\subsection{Study 2: Detecting Political Factions}
\subsubsection{Home origin distribution of top rank politicians}
According to the hometown network, we found that some provinces have larger population of politicians, some less. Combining with the ranking information, we plotted out the home origin distribution, in province level, for top ranked politicians. 

We found that all the top ranked politicians are from northeastern regions of China, and they are from regions close to the pacific east. These results are correlated to the gross domestic product distributions across the country \cite{china2017}, shown in Figure 10 ($t = -2.11, p = 0.03 $).
\begin{figure}[h]
\includegraphics[width=8cm]{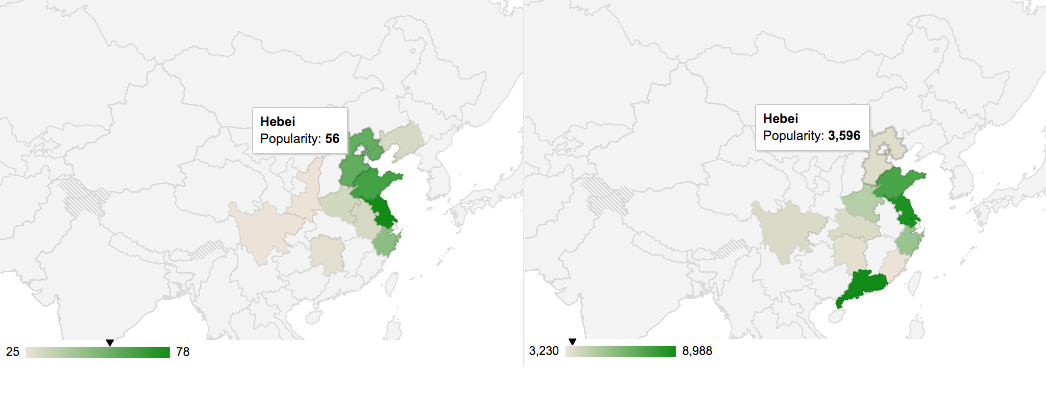}
\centering
\caption{Left is the top 10 home origin (province) distribution of the top politicians: for example, we have 56 politicians ranked in minister level from Hubei Province; Right is the gross domestic product distribution of 10 top rank provinces: for example, Hubei has 3596 dollars as its gross domestic product.}
\end{figure}
\subsubsection{Gender And Connectivity}
We calculated degree distributions for male and female candidates ($t=-2.12$, $p=0.03 $). Figure 9 showed that the average degrees of male and female politicians were close ($\mu_{male} = 341.81, \mu_{female} = 390.68$). The female politicians had higher average degrees. However, we have more high connected male politicians than female politicians. 
\begin{figure}[h]
    \centering
    \begin{subfigure}[b]{0.5\textwidth}
        \includegraphics[width=\textwidth]{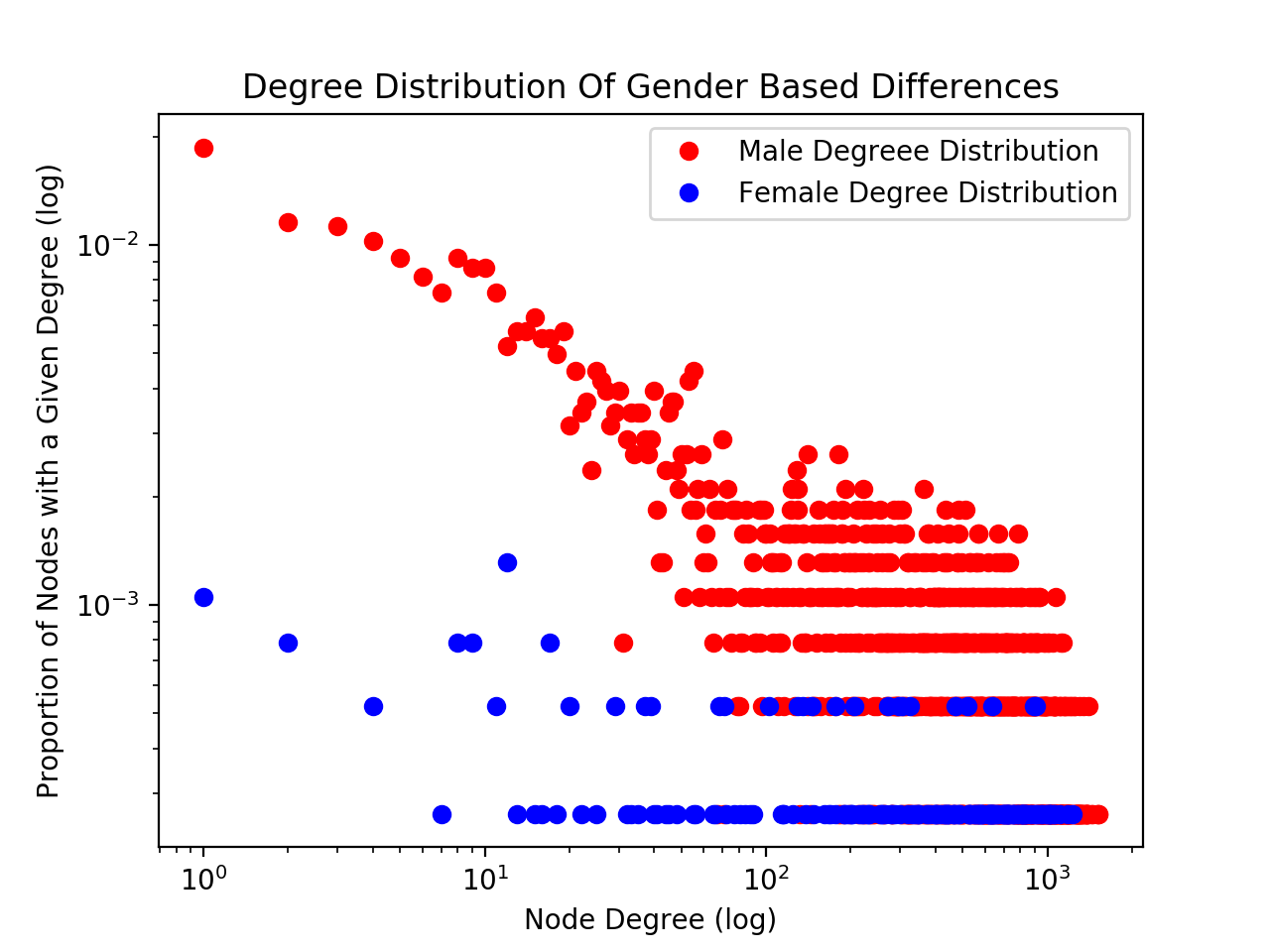}
        \label{fig:gull}
    \end{subfigure}%
    ~ 
    \begin{subfigure}[b]{0.5\textwidth}
        \includegraphics[width=\textwidth]{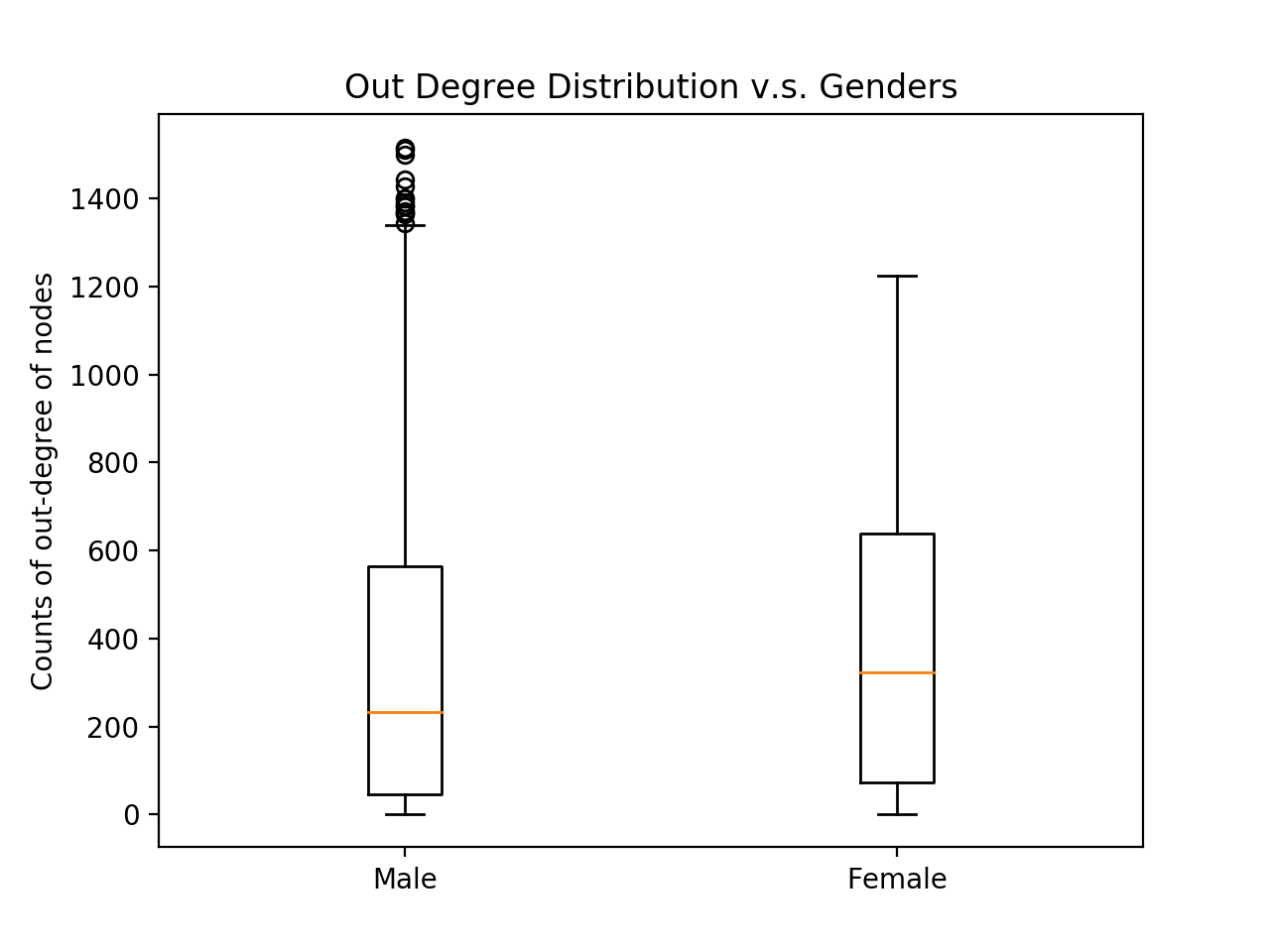}
        \label{fig:tiger}
    \end{subfigure}
    \caption{Left is the degree distributions for male and female politicians. Right is the box plot for the degree distributions.}\label{fig:animals}
\end{figure}

\subsubsection{Gender And Rankings}
\begin{figure}[h]
    \centering
    \begin{subfigure}[b]{0.5\textwidth}
        \includegraphics[width=\textwidth]{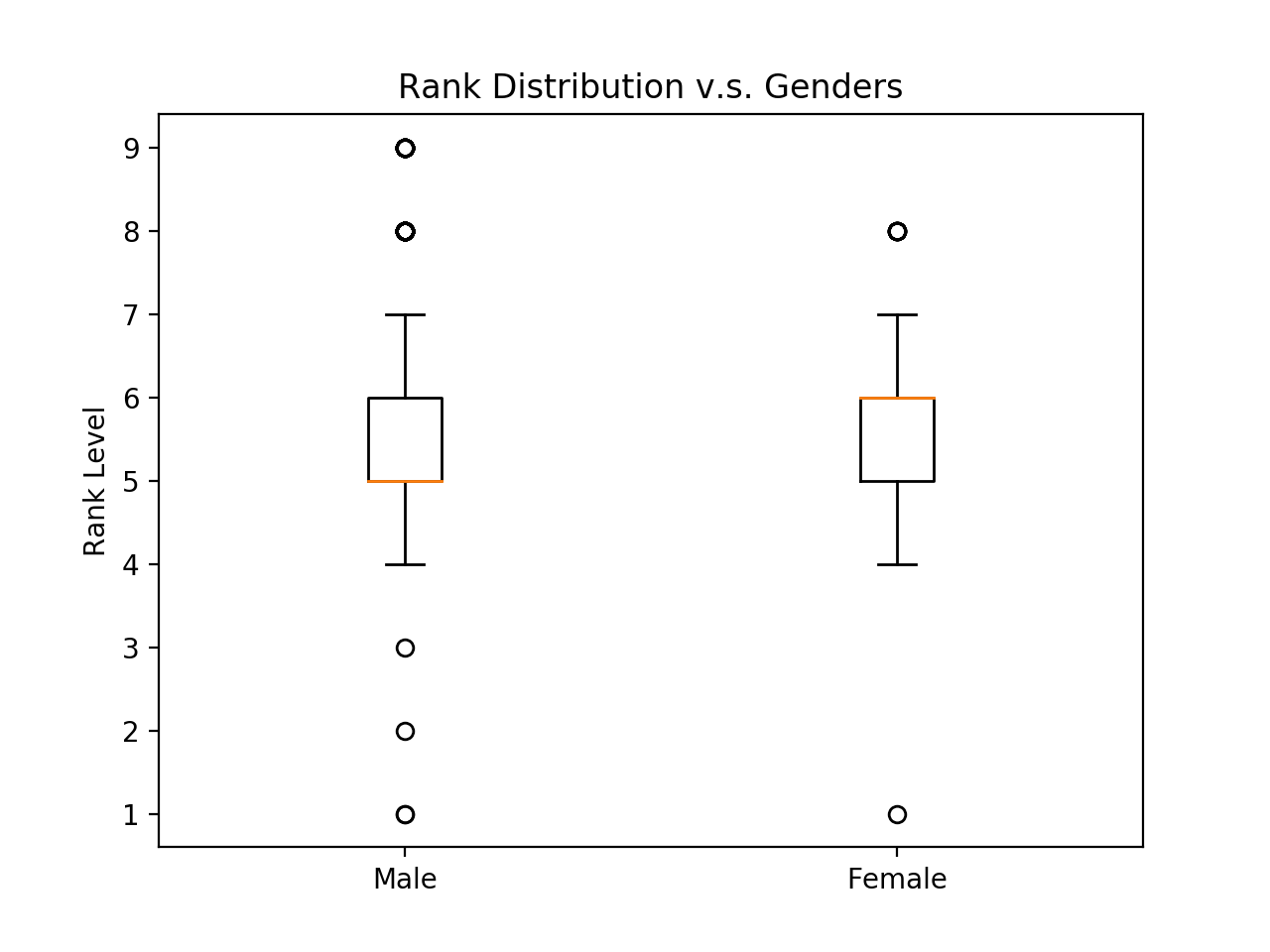}
        \label{fig:gull}
    \end{subfigure}%
    ~ 
    \begin{subfigure}[b]{0.5\textwidth}
        \includegraphics[width=\textwidth]{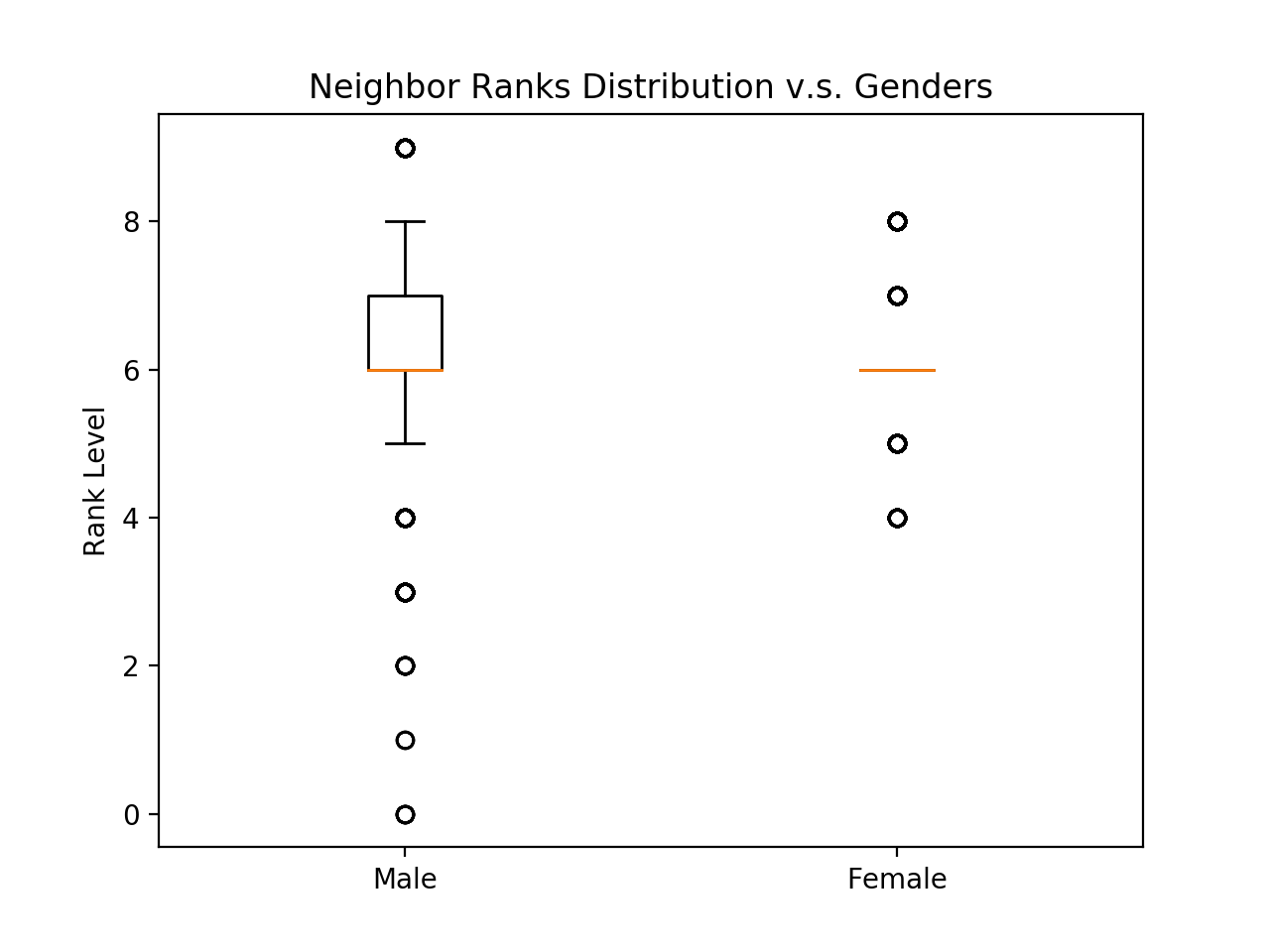}
        \label{fig:tiger}
    \end{subfigure}
    \caption{Left is the box plot for the ranking distributions for male and female politicians. Right is the box plot for the ranking distributions of neighbors for male and female politicians.}\label{fig:animals}
\end{figure}

We calculated the ranking differences between male and female politicians. Ranking distributions of male and female candidates are different. Similar to what we had for the degree distributions, the average rankings of female politicians was higher than that of male. However, we have more high ranked male politicians than female politicians.

Likewise, we calculated the average rankings for the closest neighbors of male and female candidates ($t = 3.40, p = 0.0067 $). The trend was the same.

\subsubsection{Home Origin And Similarity}
The similarity score from $node2vec$ showed politicians from the same home origin had tighter connections in general  ($t = -1.19, p = 0.24 $). 

\begin{figure}[h]
    \centering
    \begin{subfigure}[b]{0.5\textwidth}
        \includegraphics[width=\textwidth]{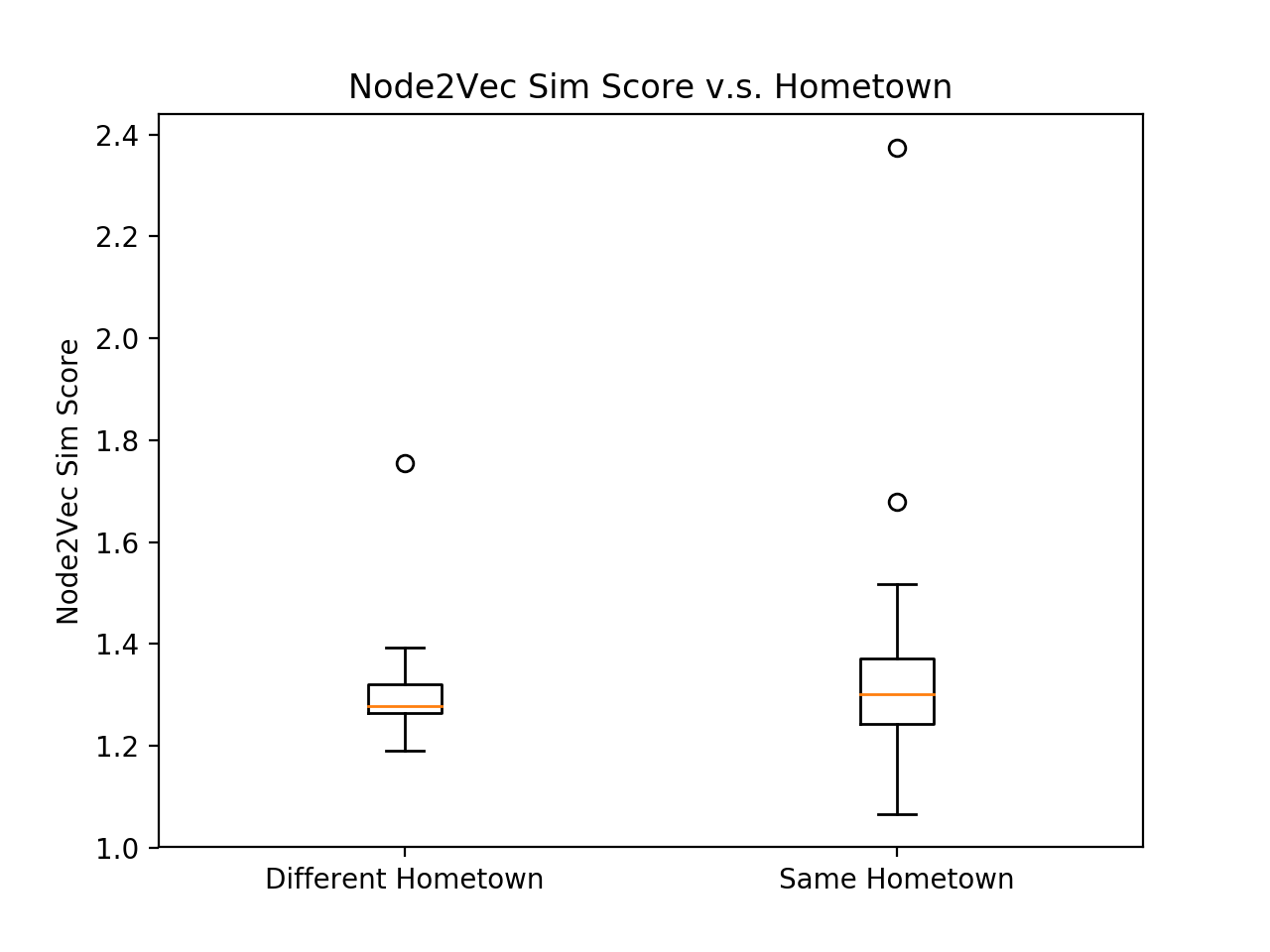}
        \label{fig:gull}
    \end{subfigure}%
    ~ 
    \begin{subfigure}[b]{0.5\textwidth}
        \includegraphics[width=\textwidth]{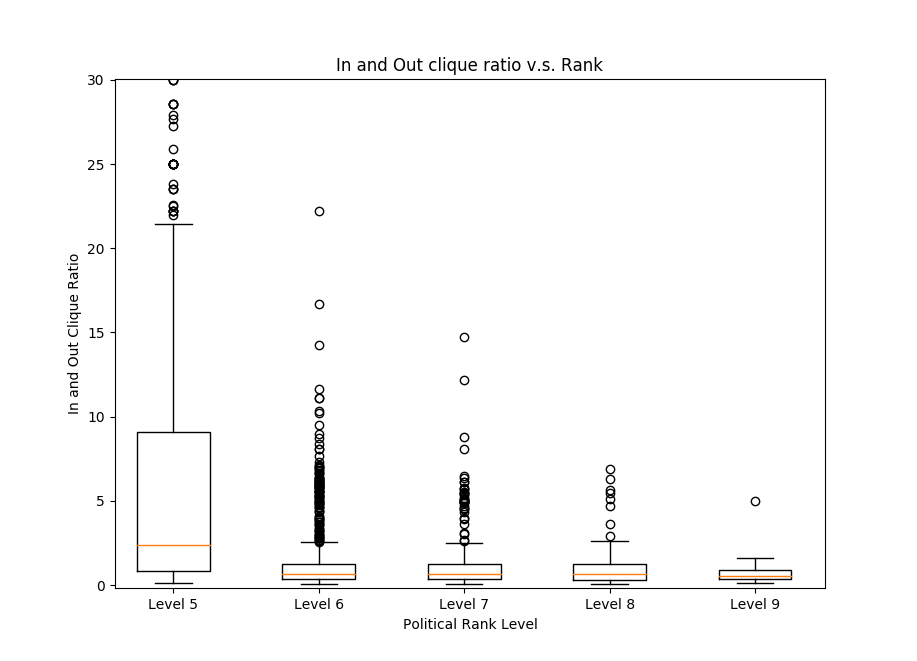}
        \label{fig:tiger}
    \end{subfigure}
    \caption{Left is the similarity score distributions for different hometown and same hometown groups. Right is the With-in clique and Inter-clique clique ratio with respect to the level rankings of the politicians. Level 9 is corresponding to the level of the president.}\label{fig:animals}
\end{figure}

In our case, hometown represented the province the politician came from. More than 70\% of the cases, politicians from the same hometown had higher connectives in the $node2vec$ embedded space. We have politicians from $Shang hai$ stood out as an outlier. The average within hometown community similarity score was $2.37$. This means politicians from $Shang hai$ tended to have stronger connections during their careers.
\subsubsection{Within Clique And Inter-clique Connectivity}
For this study, we defined that two politicians came from the same clique if they had the same home origin. Based on our (work experience) overlapping graph, we defined with-in clique connectivity as the count of edges going out from one node to other nodes who had the same home origin attribute. Similarly, we defined inter clique connectivity as the count of edges going out from one node to other nodes that had different home origin attribute. We then calculated the with-in clique and inter clique edge count ratio for each node. We plotted the ratio against the ranking level of a node. We only took nodes that had ranking level greater than 5, since they were considered as high ranking politicians. We limited the ratio from 0 to 30, since for any nodes with ratio above 30, we could consider that they only had with-in clique edges.

In the plot, we have ranking levels ranging from level 5 to level 10. Level 5 is the equivalent level for the state Governor. Level 10 is the presidential level. We can see that there is a big jump in the clique ratio from level 5 to level 6 ($t = 10.69, p \leq 0.01 $). We can see the from level 7 to level 8, there is no jump ($t = 0.71, p = 0.48 $).

\subsection{Study 3: Hubs and Authorities}
We generated four graphs based on the two networks. Shown in Figure 12, the scatter plot for the Promotion-based Network indicates that different rank groups may correspond to hubs and authorities. However, the average score of every rank does not show such relationship. Both of the hub graph and authority graph show rather flat lines for the average score, which indicates that no significant correlation is found in the Promotion-based Network.
\begin{figure}[h]
    \centering
    \begin{subfigure}[b]{0.5\textwidth}
        \includegraphics[width=\textwidth]{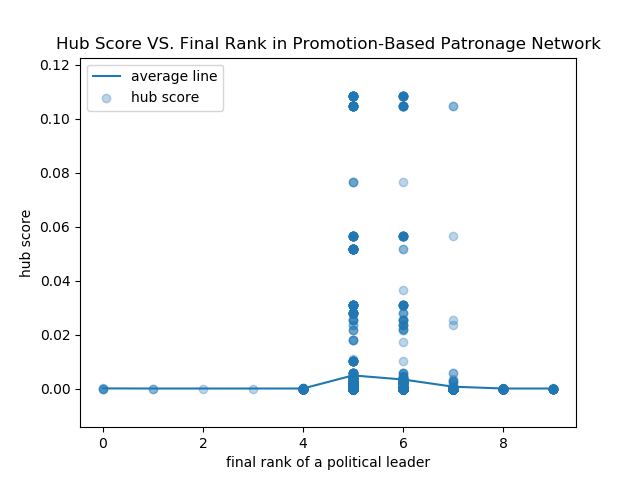}
        \label{fig:gull}
    \end{subfigure}%
    ~ 
    \begin{subfigure}[b]{0.5\textwidth}
        \includegraphics[width=\textwidth]{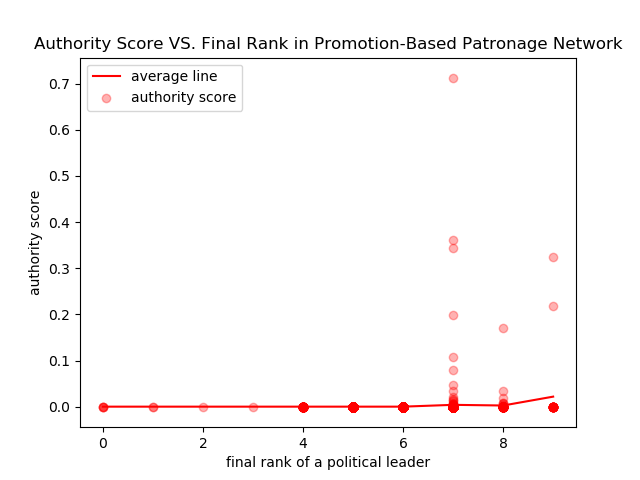}
        \label{fig:tiger}
    \end{subfigure}
    \caption{Left is the hub scores as a function of final ranks of the political leaders based on the Overlap-based Patronage Network. Right is the authority scores as a function of final ranks of the political leaders based on the Overlap-based Patronage Network.}\label{fig:animals}
\end{figure}

The Overlap-based Network, however, shows different structures. A surprising finding from the hub and authority graph is that they seem to be very similar. We explore the actual number of the scores and find that the hub and authority scores for almost every political leader are exactly the same before five decimals after the decimal point and only differ after that.
\begin{figure}[h]
    \centering
    \begin{subfigure}[b]{0.5\textwidth}
        \includegraphics[width=\textwidth]{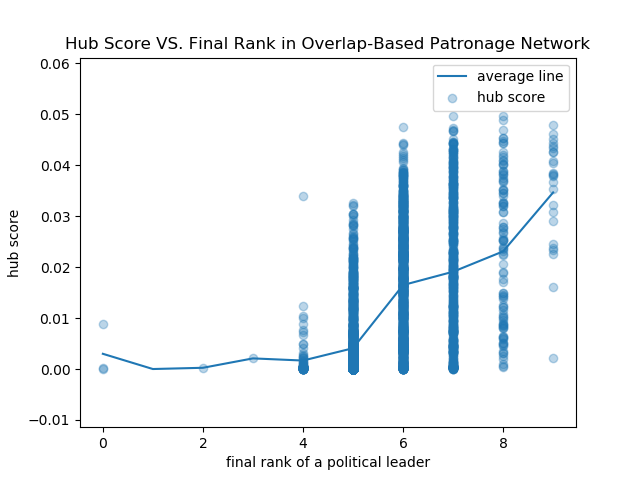}
        \label{fig:gull}
    \end{subfigure}%
    ~ 
    \begin{subfigure}[b]{0.5\textwidth}
        \includegraphics[width=\textwidth]{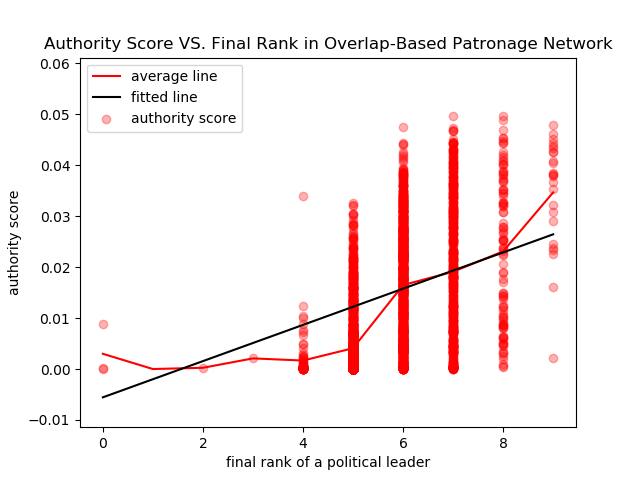}
        \label{fig:tiger}
    \end{subfigure}
    \caption{Left is the hub scores as a function of final ranks of the political leaders based on the Promotion-based Patronage Network. Right is the authority scores as a function of final ranks of the political leaders based on the Promotion-based Patronage Network.}\label{fig:animals}
\end{figure}
Based on Figure 13, both of the scatter plot and the average line show some correlation between the final rank and the hub/authority scores. Because of the similarity of both graphs, we only fit one of the graphs and believe it will be almost identical for the other one. The fitted line is shown in black on the right subgraph of Figure 13, and the statistics is shown below in the table.\\

\begin{table}[tb] \centering 
  \caption{The correlation between the authority scores and the final rank of politicians in the network.}
  \label{tab:android_intensity_increases_browser_total_time} 
  \scalebox{0.9}{
\begin{tabular}{@{\extracolsep{5pt}}lc} 
\\[-1.8ex]\hline 
\hline \\[-1.8ex] 
 & \multicolumn{1}{c}{\textit{Dependent variable:}} \\ 
\cline{2-2} 
\\[-1.8ex] & Authority Score \\ 
\hline \\[-1.8ex] 
 Final Rank Of A Political Leader & 0.00355$^{**}$ \\ 
  & (0.00064) \\ 
  (Intercept) & -0.00556$^.$ \\ 
  & (0.00342) \\ 
 \hline \\[-1.8ex] 
Observations & 10 \\ 
\hline 
\hline \\[-1.8ex] 
\textit{Note:}  & \multicolumn{1}{r}{$^{.}$p$<$0.15; $^{*}$p$<$0.05; $^{**}$p$<$0.005} \\ 
\end{tabular}}
\end{table} 
As shown in the table, there is a clear and significant correlation between the authority scores and the final rank of a political leader, indicated by the small p-value. We can say that the authority score increases as the final rank goes up for a political leader. Other statistics are reported as the following: (F(1,8) = 30.74, RSE = 0.005821, Adjusted $R^2$ = 0.7677)


\section{Discussion and Implication}
In our first study, we found that career outcome of political leaders in China is largely determined by patronage networks formed early in career, mostly even before promotion to municipality-level positions. It suggested that our network features up until 2005 adds strong prediction power to the model, implying that Chinese cadre's patronage relationships early in career may matter much more than we previously thought. 
The strong and substantial implication here is not causal and we certainly do not claim exhaustive in capturing factors determining career outcomes with around 70\% of accuracy in prediction. Nonetheless, our finding points to an interesting pattern that one's political career is largely arranged and destined by factors in her early years, be it overlapping work experience or other pre-determined features. Future research can do two things: (1) to figure out better ways of represent network features to make better predictions, e.g. Random Walks, Node Centrality, etc. Our current representation uses simple aggregation of neighborhood node basic features; (2) to explore mechanisms and factors that causally determine one's early career trajectory.

Our results from the home origin studies showed that the final ranking of a politician is highly correlated with the Gross Domestic Product (GDP) of one's home origin. This is not surprising, as we believe that higher GDP will bring a politician more resources and connections compared to a politician from a rural city in China. 

We found that the average ranking of female candidates is similar to the average rankings of male. However, there were drastically more male politicians in high-ups. This results coincided the stereotype of the gender differences in political institution. Currently, we do not have a woman among all the highest level politicians in China. These results suggested that female politician would have lower chance to get promoted in autocratic countries, such as China. We found the same gender differences in degree distribution as well.

We studied the closeness in overlap network of nodes that are from the same home origin. We found more than half of the cases, politicians with the same home origin had higher connectives in the work overlapping network. People that were from the same hometown tended to work together. We found that \textit{Shanghai} had a much higher score in this case, comparing to other cities. This suggested politicians from \textit{Shanghai} focused more on patronage relationships in political promotions.

We found politicians with higher rankings tended to have more out-clique connections. There is a big jump between level 5 and level 6. Level 5 politicians work at a state level, whereas level 6 is above state level. Our model suggested that a politician forms more out-clique connections, especially after Level 6, which makes sense in that politicians work with people from other cities and provinces after they are promoted to a national level.

In study 3, we analyzed the hub and authority scores for both the overlap-based and promotion-based networks. Our plot shows no clear correlation between the scores and final ranks on the promotion-based networks. This is due to how the promotion-based network is set up. When constructing the network, we only considered a promotee's direct superior to be his or her promoter. Higher ranked politicians are not included, which explains the lack of correlation on this graph. However, a positive linear relationship is shown on the overlap-based network. As a politician's final rank goes up, the politician tends to be more central in the network, and thus have higher HITS scores, both in terms of having more important connections and in terms of having higher authorities.

Lastly but certainly not the least, our overall analysis of patronage network information from various aspects yields a great deal of understanding for career outcome of Chinese political leaders. Again, by simply looking at the early 10-year overlapping work experience, we were able to predict career outcomes ten years later, at an accuracy level of around 70\%. This is simply amazing because the Chinese official line of cadre promotion is based on economic and governance performance, so does a major stream of literature argue. We show that patronage network, if not overwhelmingly, largely determines career ladder, and thus performance is either a second-order factor or a channel variable through which patronage factor translate itself into career outcome, i.e. politicians with better patronage relationships are assigned to regions and positions that are more likely to yield better economic performance.


\section{Limitations}

In this study, we did not exhaust the list of factors that could play a role in political promotions. In the home origin study, we only look at candidates that are from same city. The scope of a city might be too large for defining close relationships. Future studies may consider narrowing down the definition of home origin between politicians with a larger dataset.

Likewise, for the promotion network, we only considered the promotion between level 4 and level 5. Future researchers could explore the dataset to include patronage relationships incurred from every promotion during a leader's political career. Other researchers could also explore other methods in representing network structures into vectors, and conduct finer-grained research into mechanisms through which these patronage networks translate into career outcomes.


\section{Conclusion}
In this paper, we used network analysis techniques and statistical tools to understand the mechanism behind promotions in autocratic political systems. Specifically, we focused on the interpretations of the patronage network.


By using social network analysis on the three constructed networks, we found associations between the final rank of a politician and one's early stage patronage connections. We found that early stage development in the patronage network plays a more important role in one's political career compared to a more current stage. We have found the correlations between the gender and the final rankings, as well as the connectives. Our result also suggested that politicians from same hometown tend to be much closer in the patronage network. On the other hand, our results showed that in order to jump from state level position to above, one needs more out of hometown connections. Similarly, we demonstrated that hub and authority scores are direct mappings of political rankings in the patronage network. 

In conclusion, our study presented a new approach to analysis patronage relationship in autocratic political system using publicly available promotion records. Combining network analysis and statistical analysis, we were able to quantify the promotion mechanism which was long thought to be mysterious in autocratic government. We found that our results coincide with theoretical speculations from previous qualitative studies, which came from limited insider sources. We provide a new way to further quantify the promotions of autocratic political systems using publicly available data.


\bibliographystyle{spmpsci}      



\newpage
\onecolumn
\begin{center}
    \textrm{ \Large Appendix A: Level Annotation and Corresponding U.S. Positions}
\end{center}

\bigbreak
\bigbreak

\includegraphics[scale=0.4]{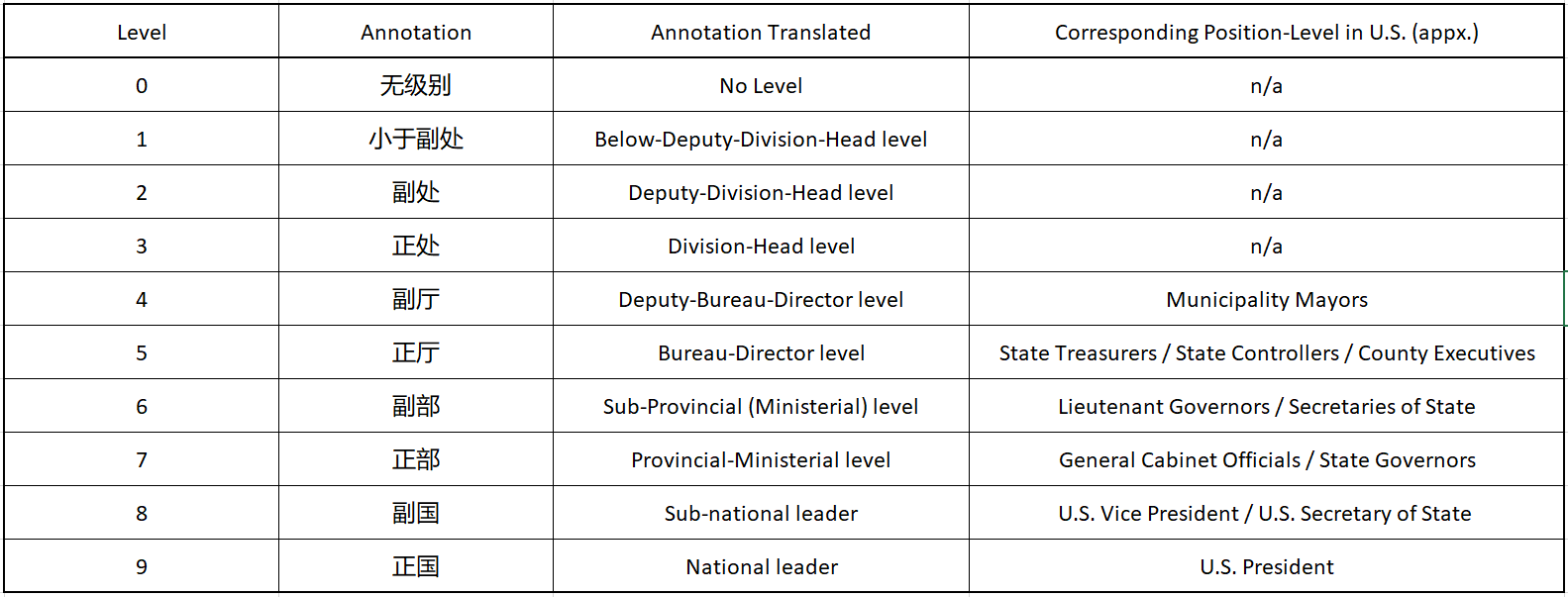}

\newpage
\begin{center}
    \textrm{\centering \Large Appendix B: Regression Table for Predicting Career Ranks}
\end{center}

\bigbreak

\includegraphics[scale=0.45]{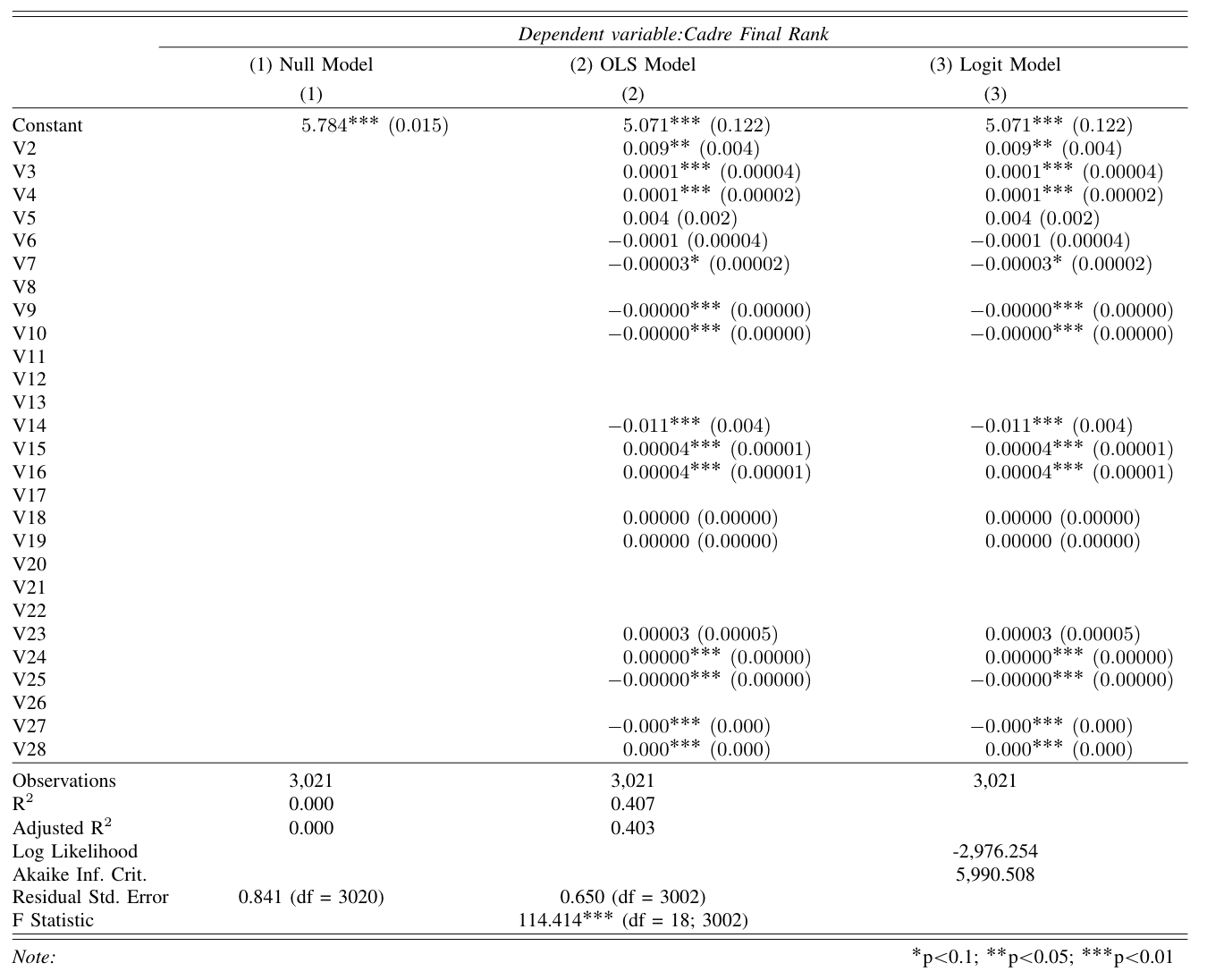}

\end{document}